\documentclass[%
preprint,
pra,
]{revtex4-2}

\usepackage{graphicx}% Include figure files
\usepackage{bm}% bold math
\usepackage{amsmath}
\usepackage{amssymb}
\usepackage{enumerate}
\usepackage{subcaption} % preambleに追加
\newcommand{\IN}{\mathrm{IN}}
\newcommand{\OUT}{\mathrm{OUT}}

\begin{document}

\preprint{}

\title{Structural Properties of the Asymmetric Barabási-Albert Model in the Lattice Limit}% Force line breaks with \\
%\thanks{}%

\author{Kazuaki Nakayama}
\email{nakayama@math.shinshu-u.ac.jp}
\affiliation{
Department of Mathematical Sciences,
Faculty of Science, Shinshu University,  
Asahi 3-1-1, Matsumoto, Nagano 390-8621, Japan
}

\author{Masato Hisakado}
\email{hisakadom@yahoo.co.jp}
\affiliation{
Nomura Holdings Inc., Otemachi 2-2-2, Chiyoda-ku, Tokyo 100-8130, Japan
}

\author{Shintaro Mori}
\email{shintaro.mori@hirosaki-u.ac.jp}
\affiliation{
Graduate school of science and Technology, 
Hirosaki University, \\
Bunkyo-cho 3, Hirosaki, Aomori 036-8561, Japan
}

\date{\today}% It is always \today, today,

\begin{abstract}
The Asymmetric BA model extends the Barabási-Albert scale-free network model by introducing a parameter $\omega$. As $\omega$ varies, the model transitions through different network structures: an extended lattice at $\omega = -1$, a random graph at $\omega = 0$, and the original scale-free network at $\omega = 1$. We derive the exact degree distribution for $\omega = -r/(r+k)$, where $k \in \{0,1,\cdots\}$, and develop a perturbative expansion around these values of $\omega$. Additionally, we show that for $\omega = -1 + \varepsilon$, the clustering coefficient scales as $\ln t / \sqrt{\varepsilon} t$ and approaches zero as $t \to \infty$, confirming the absence of small-world properties. 
\end{abstract}

\maketitle

%\tableofcontents

\section{\label{sec:intro}Introduction}
Network science has emerged as a crucial tool for understanding the structure and function of complex systems in modern society. In particular, the small-world network model proposed by Watts and Strogatz in 1998 revolutionized the modeling of many real-world networks \cite{WA1998}. This model demonstrated that networks can exhibit "small-world" properties, characterized by two critical metrics: a high clustering coefficient at the local level and short average path lengths overall. These properties allow networks to efficiently connect nodes while preserving community structures, as observed in various social, biological, and technological systems.

While small-world networks are defined by these metrics, the Barabási-Albert (BA) model, introduced by Barabási and Albert, describes a scale-free network in which the degree distribution follows a power law \cite{BA1999}. Although the BA model captures the scale-free nature of many real-world networks, it has a low clustering coefficient, meaning that it does not exhibit small-world properties \cite{BA2002}. To address this gap, several modifications of the BA model have been proposed, including the Holm-Kim model \cite{Holm2006} and the vertex deactivation model \cite{Klemm2002,Vazquez2003}, which aim to integrate both small-world properties and scale-free characteristics.

In this study, we investigate the Asymmetric BA model \cite{HM2021}, a one-parameter extension of the BA model, focusing on its structural properties when $\omega < 0$. In this model, each node has a maximum out-degree of $\lceil -r/\omega \rceil$ (using the ceiling function). Specifically, when $\omega = -1$, the out-degree of each node is $r$, and the nodes connect to their nearest neighbors, forming an extended lattice. We analyze the behavior of the clustering coefficient and average path length in the negative $\omega$ region. Additionally, we explore the conjecture that for $\omega \approx -1$, the clustering coefficient decreases asymptotically, confirming the absence of small-world properties in this regime.

This paper is structured as follows: In Section \ref{sec:model}, we recall the definition of the Asymmetric BA model. Section \ref{sec:degree} presents the derivation of the exact degree distribution for $\omega = -r/k, k\in \{r, r+1, \dots\}$. We also study the perturbative expansion of the degree distribution for $\omega = -r/k + \epsilon, k \in \{r, r+1, \dots\}$. In Section \ref{sec:cluster}, we analyze the asymptotic behavior of the clustering coefficient and average path length in the extended lattice limit $\omega = -1 + \varepsilon, \varepsilon\to 0$. Finally, we conclude with a discussion of the results and potential future research directions in Section \ref{sec:conclusion}.

\section{\label{sec:model}Asymmetric BA Model}

We recall the definition of the asymmetric Barabási-Albert (BA) model \cite{HM2021}, an evolving complex network model similar to the original BA model. The asymmetric BA model introduces two parameters: a positive integer $r$, which controls the in-degree of nodes, and a real number $\omega$ such that $\omega \geq -1$, which governs the asymmetry between the roles of out-degree and in-degree in the attachment process of new nodes.

We start with a network containing a single node, labeled as node $1$, and set the initial time as $t = 1$. At each integer time step $t$, the network consists of $t$ nodes and directed edges, forming a connected structure. A new node, labeled as node $t+1$, is added to the network, and the time variable advances by 1, so that $t := t+1$.

The rules for adding new nodes are as follows. Since the network is directed, each node $i$ (where $i = 1, 2, \dots, t$ at time $t$) is assigned an out-degree, denoted as $k^{\OUT}_{i}(t)$. The "popularity" $\ell_{i}(t)$ of each node is defined as:
\begin{equation}
  \label{eq:popularity}
  \ell_{i}(t) = \max\{0, r + \omega k_{i}^{\OUT}(t)\},\quad
  i = 1, \dots, t.
\end{equation}
Originally, the popularity was defined as $\ell_{i}(t) = k_{i}^{\IN}(t) + \omega k_{i}^{\OUT}(t)$, where $k_{i}^{\IN}(t)$ is the in-degree of node $i$ \cite{HM2021}. However, as we will demonstrate, $k_{i}^{\IN}(t) = r$ for $t \ge r+1$, making both definitions equivalent in practice. In this context, $\omega$ represents the relative weight of the out-degree compared to the in-degree. In the original BA model, $\omega = 1$, indicating symmetry between in-degree and out-degree.

Let $G_{\omega}(t)$ represent the set of nodes with positive popularity:
\begin{equation}
  \label{eq:Gomegat}
  G_{\omega}(t) = \{i \mid \ell_{i}(t) > 0,\; i = 1, \dots, t\}.
\end{equation}

There are two scenarios for adding new nodes:
(i) If the number of nodes with positive popularity, denoted as $\sharp G_{\omega}(t)$, exceeds $r$, we randomly select $r$ distinct nodes from $G_{\omega}(t)$, with selection probabilities proportional to their popularity $\ell_{i}(t)$, and create directed links from the selected nodes to the new node.
(ii) Otherwise, if $\sharp G_{\omega}(t) \leq r$, directed links are created from each node in $G_{\omega}(t)$ to the new node.

In both cases, the out-degree of any node $i$ selected from $G_{\omega}(t)$ is increased by 1:
\begin{equation}
\label{eq:increase-kout}
  k_{i}^{\OUT}(t+1) = k_{i}^{\OUT}(t) + 1,
\end{equation}
while the new node (node $t+1$) starts with an out-degree of 0:
\begin{equation}
  k_{t+1}^{\OUT}(t+1) = 0.
\end{equation}

\begin{figure}[htbp]
  \includegraphics[width=12cm]{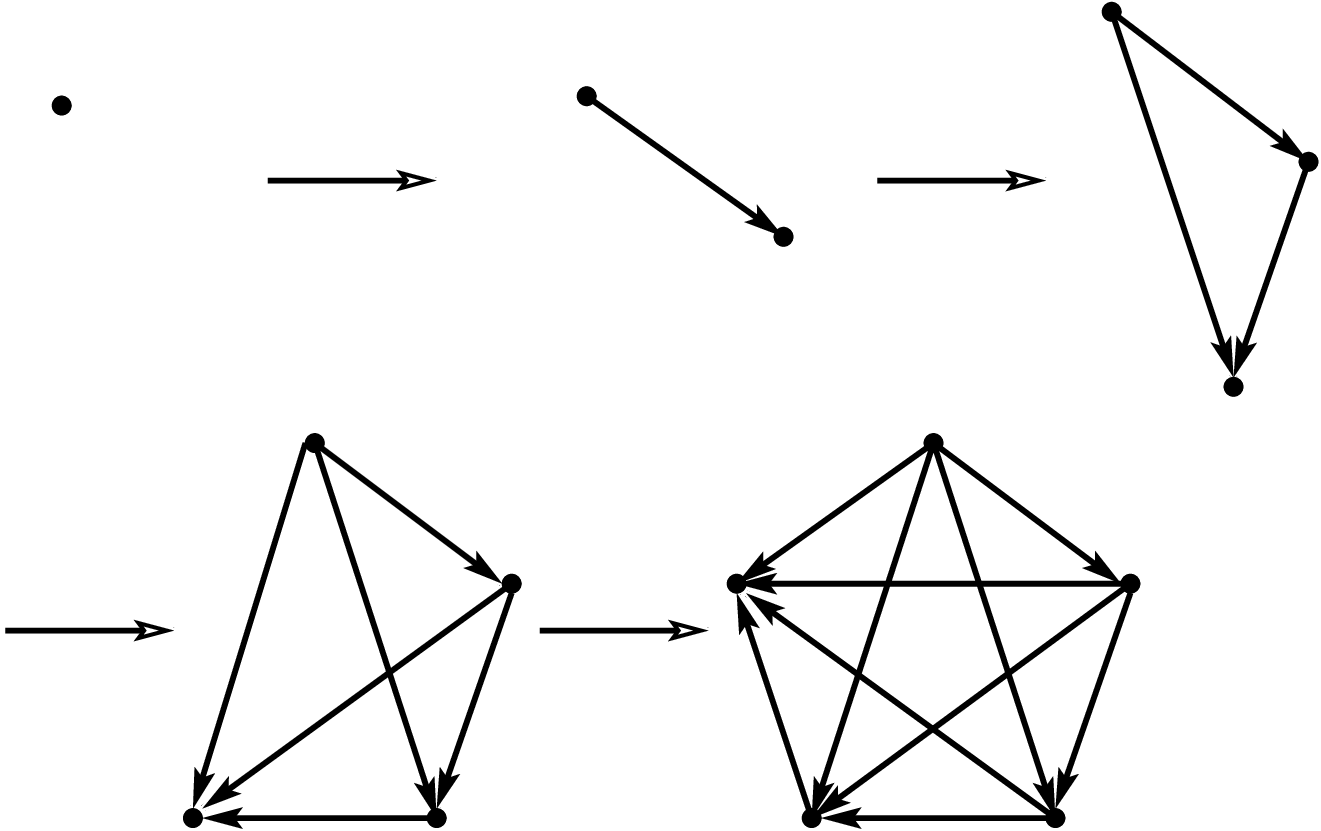}
  \caption{Time evolution of the network in its early stages for $r=4$. The network becomes fully connected at $t = r+1$.}
  \label{fig:fig1}
\end{figure}

A few important points to note:
\begin{enumerate}[(1)]
\item The network’s evolution is deterministic when $t \leq r$, and the network is fully connected when $t \leq r+1$. See Fig.~\ref{fig:fig1} for an illustration of the early stages of network evolution for $r = 4$.

\item For $t \geq r$, the number of nodes with positive popularity, $\sharp G_{\omega}(t)$, is always at least $r$ (see Appendix \ref{sec:proofofG}). 
Consequently, the in-degree of nodes with $i \geq r+1$ remains exactly $r$.
\end{enumerate}

\begin{figure}[htbp]
  \includegraphics[width=16cm]{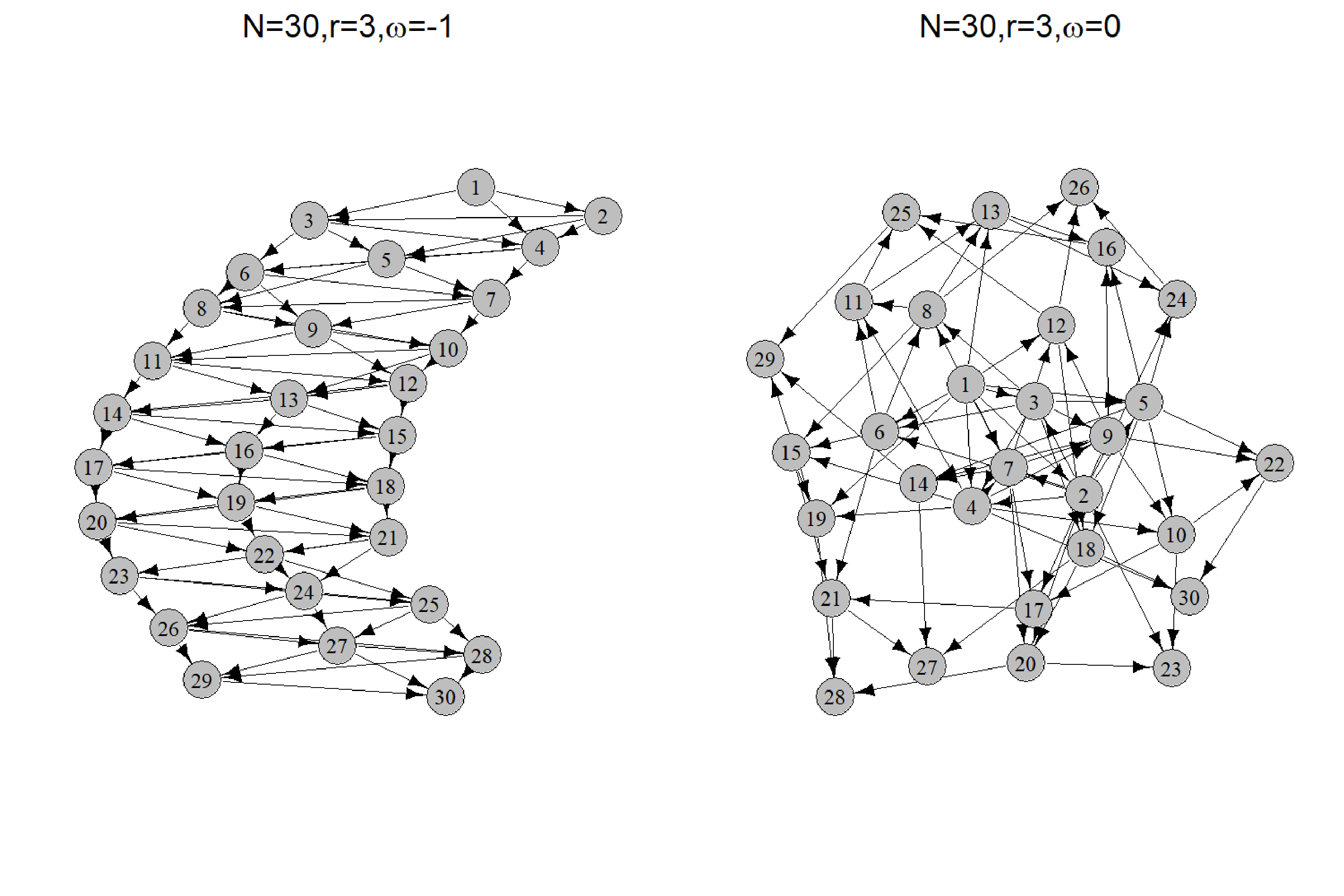}
  \caption{Examples of network configurations for $N = 30$, $r = 3$. The left panel shows the case $\omega = -1$, where the network forms an extended lattice, while the right panel shows $\omega = 0$, which corresponds to a random network.}
  \label{fig:fig2}
\end{figure}

Samples of the network are shown in Fig.~\ref{fig:fig2}. The left panel illustrates the extended lattice structure for $N = 30$, $r = 3$, and $\omega = -1$. The right panel depicts the random network configuration for the same $N$ and $r$ values, but with $\omega = 0$. In this paper, we focus on analyzing the network structure for $\omega \in (-1, 0)$.

\section{\label{sec:degree}Time Evolution of Degree Distribution}

We define the popularity function as
\begin{equation}
  \ell(k) = r + \omega k,
\end{equation}
where $r$ is a constant representing the base popularity, and $\omega$ controls the influence of the out-degree $k$. The popularity of node $i$ at time $t$, denoted by $\ell_{i}(t)$, is given by $\ell(k_{i}^{\OUT}(t))$, where $k_{i}^{\OUT}(t)$ is the out-degree of node $i$ at time $t$.

Next, we introduce the sum of the popularities for all nodes with out-degree $k$ at time $t$:
\begin{equation}
  L_{k}(t) = \sum_{i \in I_{k}(t)} \ell_{i}(t) = N_{k}(t) \ell(k),
\end{equation}
where $I_{k}(t)$ represents the set of nodes with out-degree $k$, and $N_{k}(t) = \sharp I_{k}(t)$ is the number of such nodes at time $t$. Formally, we define:
\begin{align}
  I_{k}(t) &= \{i \mid k_{i}^{\OUT}(t) = k, \; i \in G_{\omega}(t)\} \label{eq:Ikt}, \\
  N_{k}(t) &= \sharp I_{k}(t).
\end{align}

The index $k$ ranges as $0 \leq k < k_{\max} + 1$, where:
\begin{equation}
  \label{eq:kmax}
  k_{\max} =
  \begin{cases}
    \infty, & \omega \geq 0, \\
    \lceil r / (-\omega) \rceil, & \omega < 0,
  \end{cases}
\end{equation}
and $\lceil x \rceil$ is the ceiling function, representing the smallest integer greater than or equal to $x$. Additionally, note that $I_{k}(t) = \emptyset$ 
and $N_{k}(t) = 0$ for $k \geq t$, since the out-degree $k_{i}^{\OUT}(t)$ is always less than $t$.

As previously mentioned, the time evolution of the network is deterministic when $t \leq r$. Thus, we focus on the case where $t > r$. Nodes in the set $I_{k}(t)$ accept new links with probability
\begin{equation}
  \label{eq:qkt}
  q_{k}(t) = \frac{L_{k}(t)}{D(t)},
\end{equation}
where
\begin{equation}
  \label{eq:dt}
  D(t) = \sum_{k=0}^{\min\{k_{\max},t\}-1} L_{k}(t).
\end{equation}
This represents the probability of a node transitioning from the set $I_{k}(t)$ to $I_{k+1}(t+1)$.

The time evolution of the set $\{N_{k}(t)\}$ is given by:
\begin{equation}
  \label{eq:Nk}
  N_{k}(t+1) = N_{k}(t) + (q_{k-1}(t) - q_{k}(t))r,
\end{equation}
where the following conventions are adopted:
\begin{equation}
  \label{eq:convention}
  q_{-1}(t) = \frac{1}{r}, \quad
  q_{\min\{k_{\max}, t\}}(t) = 0.
\end{equation}

For the initial condition, we use the following result, obtained after the deterministic time evolution of the network:
\begin{equation}
  N_{0}(r+1) = \cdots = N_{r}(r+1) = 1,\quad
  N_{r+1}(r+1) = 0.
\end{equation}

It is straightforward to derive the conservation law:
\begin{equation}
  \sum_{k=0}^{\min\{k_{\max},t\}} N_{k}(t) = t.
\end{equation}
Thus, the distribution of out-degrees $\{p_{k}(t)\}$ is expressed as:
\begin{equation}
  \label{eq:pkt}
  p_{k}(t) = \frac{N_{k}(t)}{t},\quad
  0 \leq k \leq \min\{k_{\max}, t\}.
\end{equation}

\subsection{Degree Distribution for $-1<\omega<0$}

We focus on the case where the parameter $\omega$ is negative, 
specifically $-1 < \omega < 0$. The case $\omega = -1$ is excluded from consideration, 
as the time evolution of the network is deterministic for all $t$ (see Appendix \ref{sec:omega-1} for details).

\begin{figure}[htbp]
  \includegraphics[width=12cm]{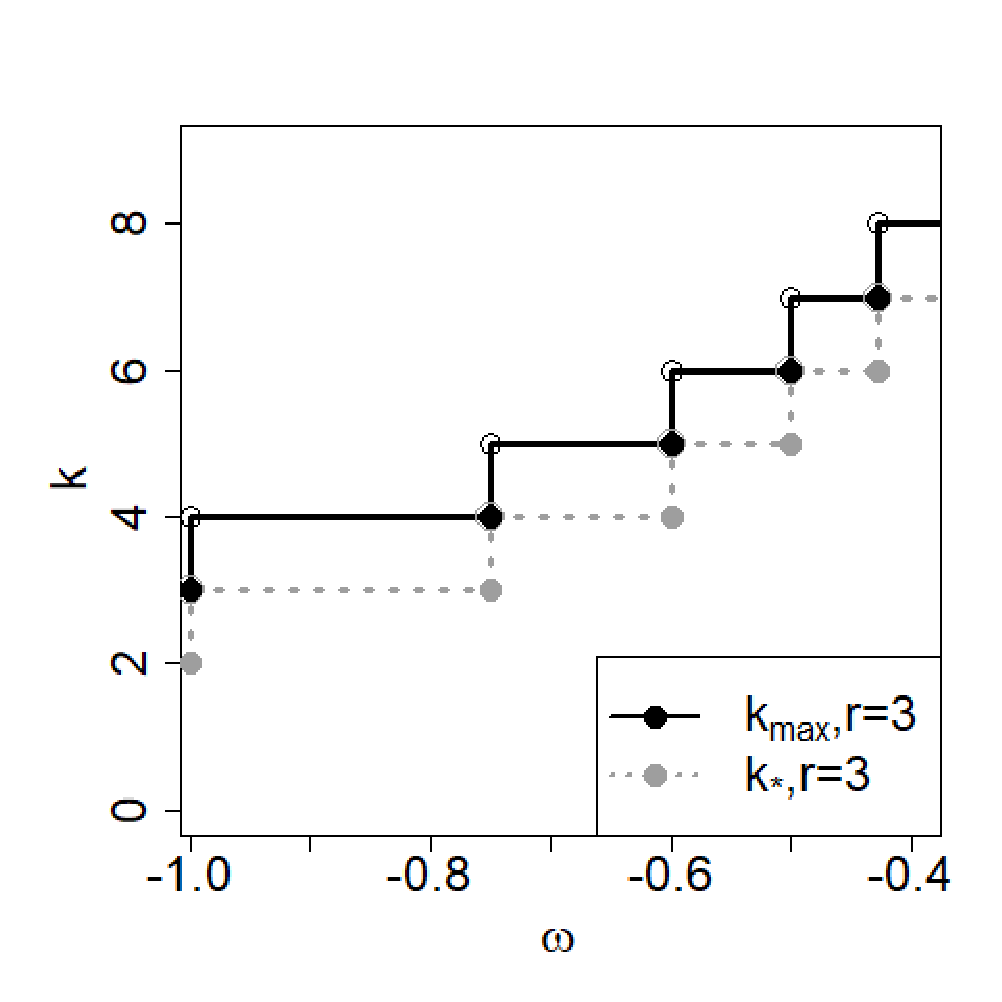}
  \caption{Plot of $k_{\max}$ and $k_{\ast}$ vs. $\omega\in [-1,-0.4]$ for $r=3$.}
  \label{fig:fig3}
\end{figure}

From eq.~\eqref{eq:kmax}, we observe that $k_{\max}$ is finite and greater than $r$. The condition that the out-degree $k$ lies in the range $0 \leq k \leq k_{\max}$ for some $k_{\max} > r$ is expressed as:
\begin{equation}
  -\frac{r}{k_{\ast}} < \omega \leq -\frac{r}{k_{\max}},
\end{equation}
where we introduce the simplified notation:
\begin{equation}
  k_{\ast} = k_{\max} - 1.
\end{equation}
Fig.~\ref{fig:fig3} shows the plot of $k_{\max}$ and $k_{\ast}$ as functions of $\omega$.
When $\omega\in (r/(k-1),r/k],k\in\{r+1,\cdots\}$, $k_{max}=k$.

Next, we assume the network has reached a steady state, meaning:
\begin{equation}
  q_{k}(t) = q_{k} \quad (\text{constant}).
\end{equation}
With this assumption, the recursion relation \eqref{eq:Nk} can be solved. 
The solution is:
\begin{align}
  N_{k}(t) &=
  \begin{cases}
    \theta(k < t), & t \leq r, \\
    (t - r - 1)(q_{k-1} - q_{k})r + \theta(k \leq r), & t > r,
  \end{cases}
  \notag \\
  &\simeq t(q_{k-1} - q_{k})r, \quad \text{when } t \gg r,
  \label{eq:Nkt}
\end{align}
where we have used the Boolean function defined as:
\begin{equation}
  \theta(P) =
  \begin{cases}
    1, & \text{if the logical statement $P$ is true,} \\
    0, & \text{if the logical statement $P$ is false.}
  \end{cases}
\end{equation}
Considering the definition of $q_{k}$ in eq.~\eqref{eq:qkt}, we introduce the following ratio:
\begin{align}
  u_{k} &= \frac{N_{k}(t)}{\sum_{k=0}^{k_{\ast}} N_{k}(t)}, \quad 0 \leq k \leq k_{\ast}, 
  \notag \\
  &\simeq R(q_{k-1} - q_{k}), \quad R = \frac{r}{1 - q_{k_{\ast}}r}.
  \label{eq:uk}
\end{align}
Then, the transition probability $q_{k}$ is given by:
\begin{align}
  \label{eq:qk}
  q_{k} &\simeq \frac{u_{k} \ell(k)}{A},
  \\
  A &= R \{1 + \omega - \ell(k_{\max}) q_{k_{\ast}}\}.
  \label{eq:A}
\end{align}
Note that $D(t)$, defined in eq.~\eqref{eq:dt}, can be expressed as:
\begin{equation}
  \label{eq:Dt}
  D(t) \simeq \frac{A}{R} rt.
\end{equation}
By eliminating $u_{k}$ from eqs.~\eqref{eq:uk} and \eqref{eq:qk}, we obtain the following recursion relation:
\begin{equation}
  \label{eq:Bk}
  q_{k-1} = B_{k} q_{k}, \quad
  B_{k} = 1 + \frac{A/R}{\ell(k)}.
\end{equation}
Thus, we conclude that the probability $q_{k_{\ast}}$ of the final transition 
$I_{k_{\ast}} \rightarrow I_{k_{\max}}$ is the solution of the following 
algebraic equation of degree $k_{\max}$:
\begin{equation}
  \label{eq:qk*}
  q_{k_{\ast}} \sum_{k=0}^{k_{\ast}} \prod_{i=k+1}^{k_{\ast}} B_{i} = 1,
\end{equation}
where we adopt the convention that the empty sum is 0 and the empty product is 1.

\paragraph*{Case 1. $\omega = -r/k_{\max},k_{max}\in \{r+1,r+2,\cdots\}$:}

In this case, eq.~\eqref{eq:qk*} is easily solved as:
\begin{equation}
  q_{k_{\ast}} = \left(\frac{k_{\max}}{r}\right) B\left(\frac{k_{\max}}{r}, k_{\max}\right),
\end{equation}
where $B(x, y)$ is the beta function \cite{AS}. Then, using eq.~\eqref{eq:Bk}, $q_{k}$ can be obtained as:
\begin{equation}
  q_{k} = \frac{q_{k_{\ast}}}{(k_{\max}/r - 1)B(k_{\max}/r - 1, k_{\max} - k)},
  \quad
  k = 0, \dots, k_{\ast}.
\end{equation}
The out-degree distribution $p_{k}$, defined in eq.~\eqref{eq:pkt}, is expressed using eqs.~\eqref{eq:Nkt} and \eqref{eq:Bk} as:
\begin{equation}
  p_{k} = r \frac{k_{\max}/r - 1}{k_{\max} - k} q_{k}, \quad
  k = 0, \dots, k_{\max}.
\end{equation}
Note that $p_{k_{\max}} = r q_{k_{\ast}}$ is finite, as $k = k_{\max}$ is a zero of $q_{k}$ of order 1.

We also note that $\partial p_{k}/\partial k \gtreqless 0$ if $k_{\max} \lesseqgtr 2r$. 
In particular, $p_{k}$ does not depend on the out-degree $k$ 
when $k_{\max} = 2r$ \cite{HM2021}.

In the limit $k_{\max} \rightarrow \infty$, i.e., as $\omega \to 0_{-}$, the out-degree distribution becomes:
\begin{equation}
  p_{k} = \frac{1}{r} \rho^{-k-1}, \quad k = 0, 1, 2, \dots,
\end{equation}
where
\begin{equation}
  \rho = 1 + \frac{1}{r}.
\end{equation}

\begin{figure}[htbp]
  \includegraphics[width=16cm]{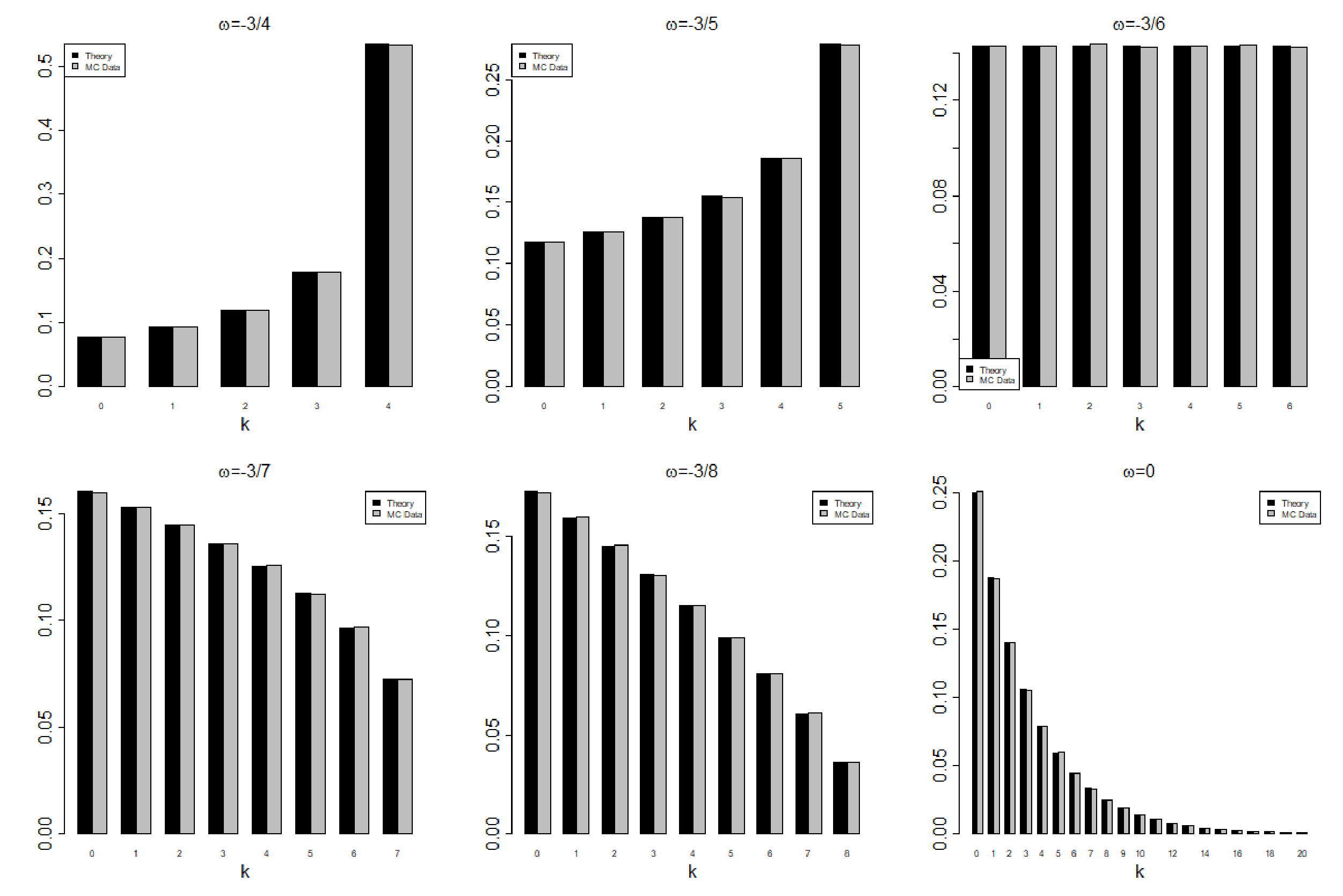}
  \caption{Plot of $p_{k}$ vs. $k$ for $r=3$ and
  $\omega \in \{-3/4, -3/5, -3/6, -3/7, -3/8, 0\}$.
  $N=10^4$, with $10^2$ samples.}
  \label{fig:fig4}
\end{figure}

Fig.~\ref{fig:fig4} shows the plot of $p_k$ vs. $k$ for $r=3$. $\omega \in \{-3/4, -3/5, -3/6, -3/7, -3/8, 0\}$ and  $k_{max}\in \{4,5,6,7,8,\infty\}$.
As can be seen clearly, the numerical results match the theoretical predictions.
The degree distribution becomes discrete uniform distribution for $k_{max}=2r=6$.
For $k<2r(>2)$, $p_{k}$ is an increasing (decreasing) function of $k$.
In the limit $k_{max}\to \infty$ and $\omega\to 0$, 
 the degree distribution becomes the geometric distribution with the success 
 probability $r/(r+1)$.

\paragraph*{Case 2. $\omega = -r/k_{\max} + \varepsilon$, $\varepsilon < 0$:}

We solve equation \eqref{eq:qk*} perturbatively:
\begin{equation}
  \label{eq:qkeps}
  q_{k_{\ast}} = q_{k_{\ast}}^{(0)} + \varepsilon q_{k_{\ast}}^{(1)} + \cdots.
\end{equation}
The zeroth-order term was determined in Case 1. 
The first-order term is expressed as:
\begin{equation}
  q_{k_{\ast}}^{(1)} = q_{k_{\ast}}^{(0)} \frac{k_{\max}^{2}}{r}
  \left\{ \chi(1, k_{\max}) - (\rho - q_{k_{\ast}}^{(0)}) \chi(k_{\max}/r, k_{\max}) \right\}.
\end{equation}
The function $\chi(x, n)$ is defined in Appendix \ref{sec:function-chi}. 
The transition probability $q_{k}$ and the out-degree distribution $p_{k}$ 
are also expanded in terms of $\varepsilon$. 
The zeroth-order terms $q_{k}^{(0)}$ and $p_{k}^{(0)}$ are the same as presented 
in Case 1. The first-order terms are:
\begin{subequations}
\begin{align}
  q_{k}^{(1)} &=
  \begin{cases}
    \begin{aligned}[b]
      &q_{k}^{(0)} \frac{k_{\max}^{2}}{r}
      \left\{\vphantom{q_{k_{\ast}}^{(0)}} \chi(k_{\max} - k, k + 1)
      \right.
      \\
      &\quad \left. - (\rho - q_{k_{\ast}}^{(0)}) \chi(\rho k_{\max} - k - 1, k + 1) \right\}
    \end{aligned},
    & 0 \leq k \leq k_{\ast},
    \\
    0, &k = k_{\max},
  \end{cases}
  \\
  p_{k}^{(1)} &= r(q_{k-1}^{(1)} - q_{k}^{(1)}), \quad 0 \leq k \leq k_{\max}.
\end{align}
\end{subequations}
Note that $q_{-1}^{(1)} = 0$.

\begin{figure}[htbp]
  \includegraphics[width=15cm]{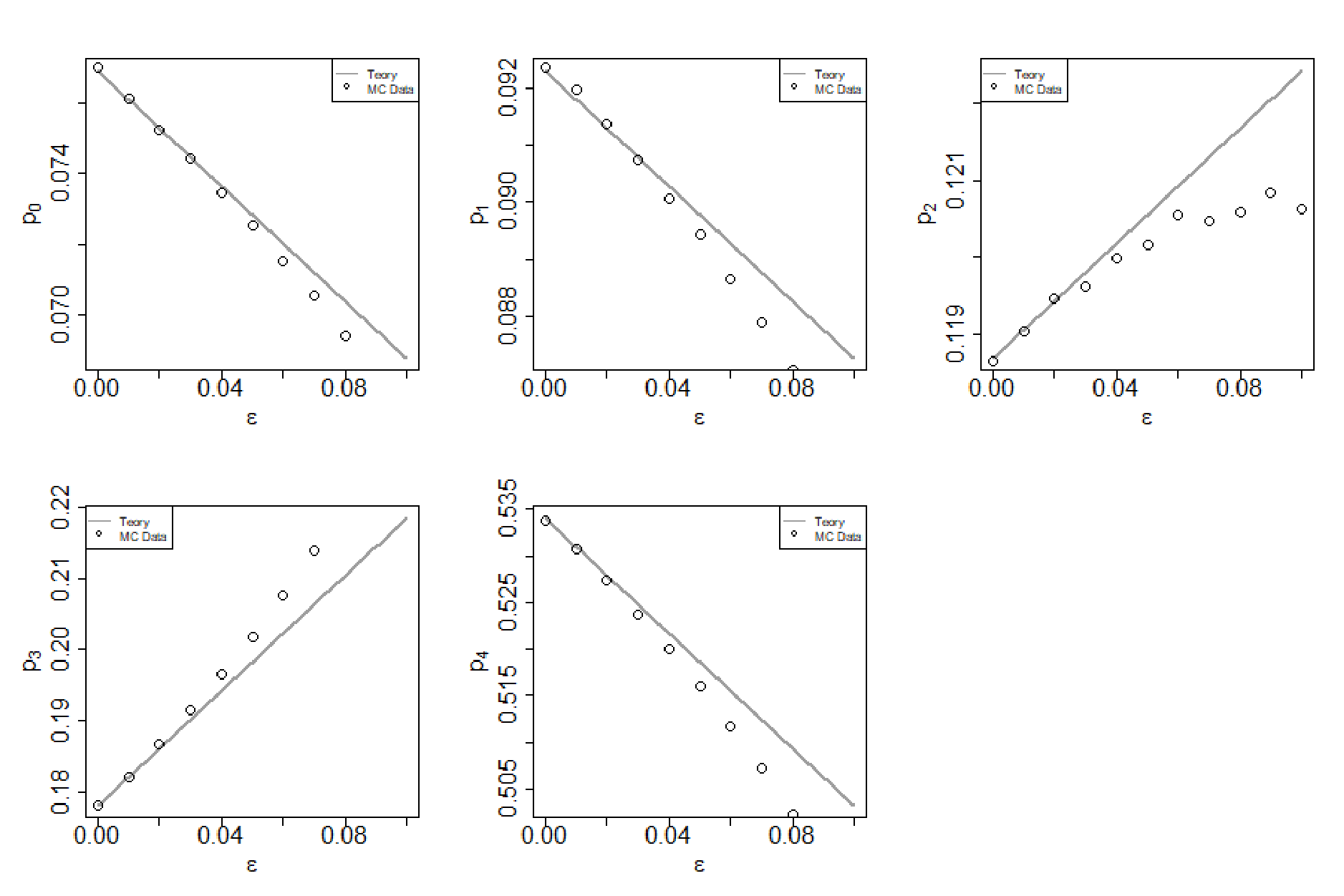}
  \caption{Plot of $p_{k}, k \in \{0,1,2,3,4,5\}$ vs. $\varepsilon$ for $r=3$, $k_{max}=4$.
  $\omega=-3/4-\varepsilon$ and $\varepsilon \in \{0.01,0.02,\cdots,0.1\}$. 
  $N=10^4$, with $10^2$ samples.}
  \label{fig:fig5}
\end{figure}

Fig.~\ref{fig:fig5} shows the plot of $p_k$ vs. $\varepsilon$ for $r=3$ and $k_{max}=4$.
$\omega=-3/4-\varepsilon$ and $\varepsilon \in \{0.01,0.02,\cdots,0.1\}$.
The solid lines represent the results of the first-order perturbative calculation in $\varepsilon$. The plots of the Monte Carlo (MC) data closely align with them, confirming the theoretical predictions.

\paragraph*{Case 3. $\omega = -r/(k_{max}-1) + \varepsilon$, $k_{max} > r+1$, $\varepsilon > 0$:}

As in the case 2, we solve eq.~\eqref{eq:qkeps} pertubatively. 
The zeroth-order terms are essentially the same as those in Case 1:
\begin{subequations}
\begin{align}
  q_{k}^{(0)} &=
  \frac{(k_{\ast}/r) B(k_{\ast}/r, k_{\ast})}{(k_{\ast}/r - 1) B(k_{\ast}/r - 1, k_{\ast} - k)},
  \quad
  0 \leq k \leq k_{\ast},
  \\
  p_{k}^{(0)} &=
  \begin{cases}
    r \dfrac{k_{\ast}/r - 1}{k_{\ast} - k} q_{k}^{(0)}, & 0 \leq k \leq k_{\ast}, \\
    0, & k = k_{\max}.
  \end{cases}
\end{align}
\end{subequations}

The first-order terms differ from those in Case 2. We have:
\begin{subequations}
\begin{align}
  q_{k}^{(1)} &=
  \begin{cases}
    \begin{aligned}[b]
      &q_{k}^{(0)} \frac{k_{\ast}^{2}}{r} \left\{
      \vphantom{\left(\rho + \frac{q_{k_{\ast}}^{(1)} r}{k_{\ast}^{2}}\right)}
        \chi(k_{\ast} - k, k + 1)
      \right.
      \\
      &\quad\left.
        - \left(\rho + \frac{q_{k_{\ast}}^{(1)} r}{k_{\ast}^{2}}\right)
        \chi(\rho k_{\ast} - k - 1, k + 1)
      \right\}
    \end{aligned},
    & 0 \leq k \leq k_{\ast},
    \\
    0, & k = k_{\max},
  \end{cases}
  \\
  p_{k}^{(1)} &= r (q_{k-1}^{(1)} - q_{k}^{(1)}), \quad 0 \leq k \leq k_{\max}.
\end{align}
\end{subequations}

Note that $q_{-1}^{(1)} = 0$ and $q_{k_{\ast}}^{(1)} = (k_{\ast}/r)^{2} B(k_{\ast}/r - 1, k_{\max})$. The number of nodes in $I_{k_{\max}}(t)$ is of first order: $N_{k_{\max}}(t) = \varepsilon p_{k}^{(1)} t$.

\begin{figure}[htbp]
  \includegraphics[width=15cm]{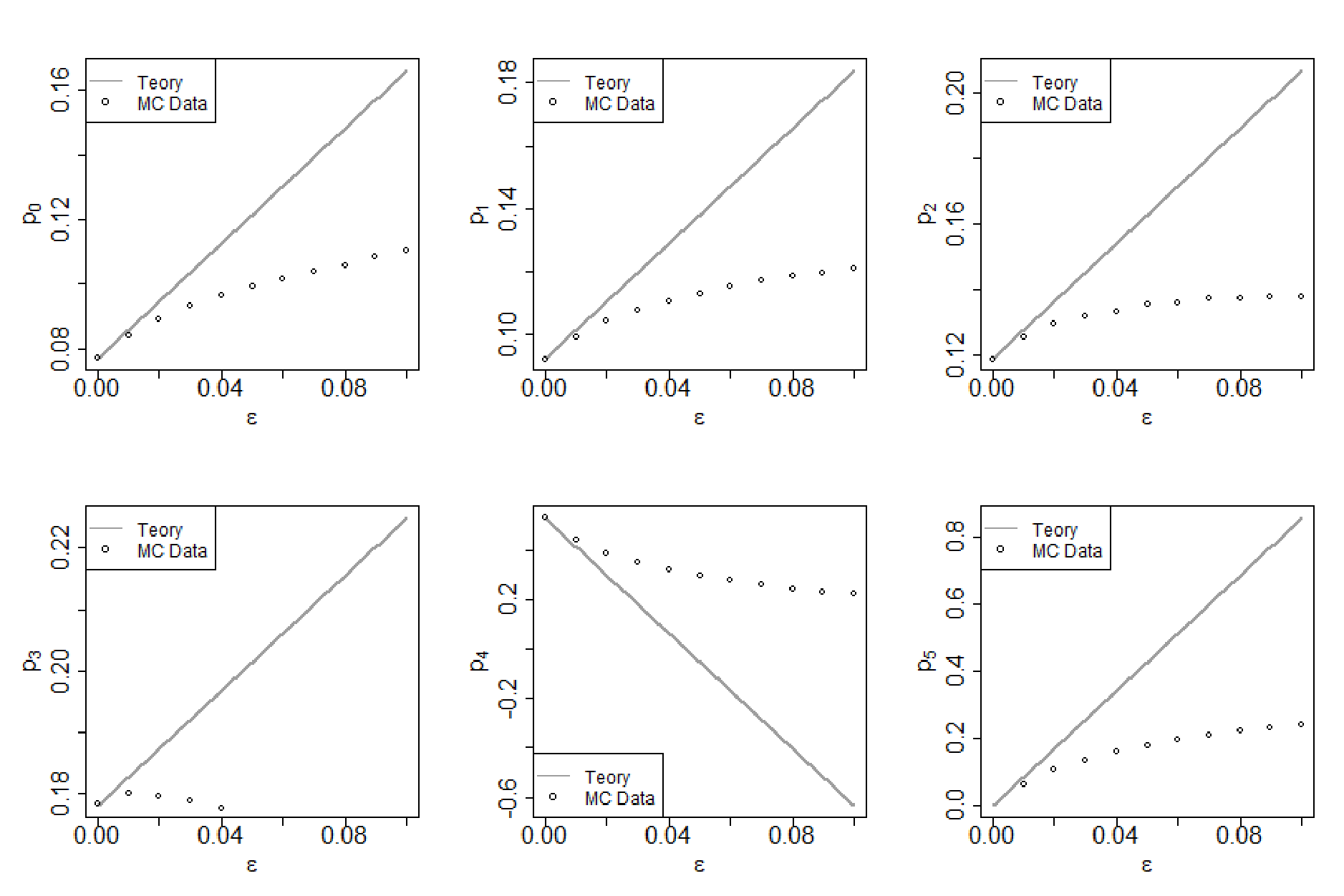}
  \caption{Plot of $p_{k},k\in \{0,1,2,3,4,5\}$ vs. $\varepsilon$ for $r=3,k_{Max}=5$.
   $\omega=-3/4+\varepsilon$ and $\varepsilon \in \{0.01,0.02,\cdots,0.1\}$.
  $N=10^4$, with $10^2$ samples.}
  \label{fig:fig6}
\end{figure}

Fig.~\ref{fig:fig6} shows the plot of $p_k$ vs. $\varepsilon$ for $r=3$ and $k_{max}=5$.
$\omega=-3/4+\varepsilon$ and $\varepsilon \in \{0.01,0.02,\cdots,0.1\}$.
The solid lines represent the results of the first-order perturbative calculation in $\varepsilon$. The plots of the Monte Carlo (MC) data closely align with them, confirming the theoretical predictions.

\paragraph*{Case 4. $\omega = -1 + \varepsilon$,$k_{max}=r+1$,$\varepsilon > 0$}

The perturbative expansion in this case is fundamentally 
different from eq.~\eqref{eq:qkeps}, as it includes a $\varepsilon^{1/2}$ term (see Appendix \ref{sec:expansion}):
\begin{equation}
  q_{k} = q_{k}^{(0)} + \varepsilon^{1/2} q_{k}^{(1/2)} + \varepsilon q_{k}^{(1)} + \cdots.
\end{equation}
Our result for the transition probabilities are:
\begin{subequations}
\begin{align}
  q_{k}^{(0)} &= \frac{1 - \delta_{kr}}{r}, \\
  q_{k}^{(1/2)} &= \delta_{kr} - \frac{1 - \delta_{kr}}{r} \chi(r - k, k + 1),
\end{align}
\end{subequations}
for $0 \leq k \leq r$. The out-degree distribution $p_{k}$ is expressed as:
\begin{subequations}
\begin{align}
  p_{k}^{(0)} &= \delta_{kr}, \\
  p_{k}^{(1/2)} &= 
  \begin{cases}
    \frac{1}{r - k}, & 0 \leq k < r, \\
    -(r + \chi(1, r)), & k = r, \\
    r, & k = r + 1,
  \end{cases}
\end{align}
\end{subequations}
where $\delta_{kr}$ is the Kronecker delta. Note that:
\begin{equation}
  p_{r} \sim O(1), \quad p_{k} \sim O(\varepsilon^{1/2}), \quad k \neq r.
\end{equation}

\begin{figure}[htbp]
  \includegraphics[width=10cm]{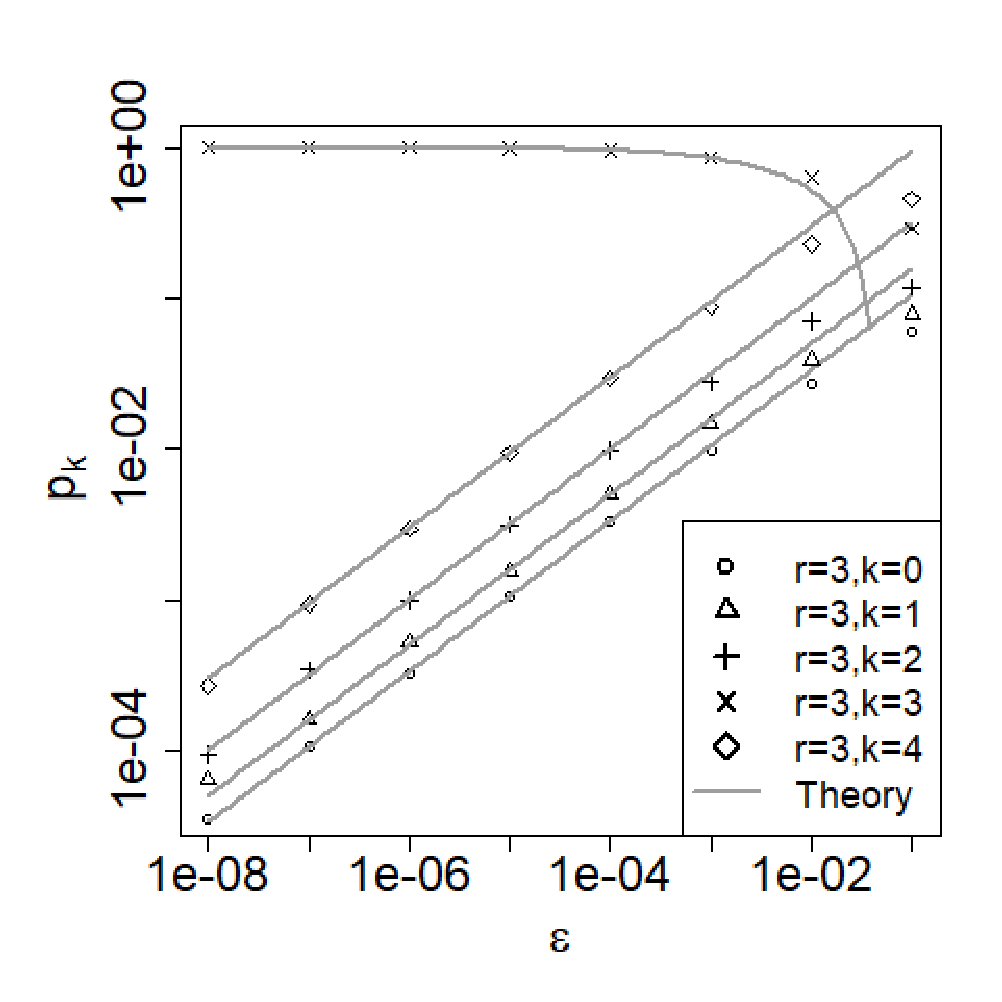}
  \caption{Plot of $p_{k},k\in \{0,1,2,3,4\}$ vs. $\varepsilon$ for $r=3,k_{max}=4$.
  $\omega=-1+\varepsilon$ and $\varepsilon \in \{10^{-8},10^{-7},\cdots,10^{-1}\}$.
  $N=10^5$, with $10^1$ samples.}
  \label{fig:fig7}
\end{figure}

Fig.~\ref{fig:fig7} shows the double logarithmic plot of $p_k$ vs. $\varepsilon$ for $r=3$ and $k_{max}=4$. $\omega=-1+\varepsilon$ and $\varepsilon \in \{10^{-8},10^{-7},\cdots,10^{-1}\}$.
The gray solid lines represent the results of the perturbative calculation in $\varepsilon^{1/2}$. The plots of the Monte Carlo (MC) data closely align with them, confirming the theoretical predictions.

\section{\label{sec:cluster}Cluster Coefficient}

In this section, we derive the asymptotic behavior of the cluster coefficient.

\subsection{Popularity Dynamics}

According to the time evolution rule, $\ell_{i}(t)$ evolves as:
\begin{equation}
  \ell_{i}(t+1) = \ell_{i}(t) +
  \begin{cases}
    \omega, & \text{with probability}\; r\beta\ell_{i}(t)/D(t), \\
    0, & \text{with probability}\; 1 - r\beta\ell_{i}(t)/D(t),
  \end{cases}
\end{equation}
where $\beta \lesssim 1$ is a parameter to be determined shortly. In the continuous time limit, this leads to the following stochastic differential equation \cite{G2009}:
\begin{equation}
  d\ell_{i}(t)
  = \omega \frac{r\beta\ell_{i}(t)}{D(t)} dt
  + |\omega| \sqrt{\frac{r\beta\ell_{i}(t)}{D(t)} \left(1 - \frac{r\beta\ell_{i}(t)}{D(t)} \right)} dW_{t},
\end{equation}
where $W_{t}$ is the Wiener process. Ignoring fluctuations, we obtain the classical solution:
\begin{equation}
  \ell_{i}(t) = r \exp\left( \int_{i}^{t} \frac{r\beta\omega}{D(t)} dt \right), \quad t \geq i.
\end{equation}
We also treat the variable $i$ as continuous. Consistency requires that:
\begin{equation}
  D(t) = \int_{0}^{t} \ell_{i}(t) \, di,
\end{equation}
leading to the conclusion:
\begin{align}
  \label{eq:beta}
  \beta &= 1 - \frac{\ell(k_{\max}) p_{k_{\ast}}}{\omega}, \\
  \ell_{i}(t) &= r \left( \frac{t}{i} \right)^{\frac{\beta\omega}{1 + \beta\omega}}, \quad t \geq i.
\end{align}
Note that $1 \geq \beta > 1 - p_{k_{\ast}}$.

\subsection{Cluster Coefficient}

Let $i$ and $j$ be two nodes such that $i<j$.
The probability of existence of a directed link
$i\rightarrow j$ is given by
\begin{equation}
  p_{ij} = r\beta\frac{\ell_{i}(j-1)}{D(j-1)}.
\end{equation}
Since the direction of edge is not essential, we extend $p_{ij}$
to a symmetric adjacency matrix,
\begin{equation}
  p_{ij} =
  \begin{cases}
    p_{ji}, & i>j,
    \\
    0, & i=j.
  \end{cases}
\end{equation}
The cluster coefficient $C(t)$ is expressed as
\begin{align}
  C(t) &= \frac{1}{t}\sum_{i=1}^{t}C_{i}(t).
  \\
  C_{i}(t) &= \frac{2}{k_{i}(k_{i}-1)}\sum_{1\le j<k\le t\atop
  j,k\neq i}p_{ij}p_{ik}p_{jk},
\end{align}
where
$k_{i}=k_{i}(t)$ is the degree of the node $i$ at time $t$.
$C_{i}(t)$ is the cluster coefficient of the node $i$.

To estimate $C(t)$, we replace $k_{i}$ by its mean value $(1+\beta)r$ and 
adopt the continuous limit:
\begin{align}
  C(t) &\simeq \frac{6}{(1+\beta)^{2}r^{2}t} \iiint_{0 \le i < j < k \le t} p_{ij} p_{ik} p_{jk} \, di \, dj \, dk, \\
  p_{ij} &= r \beta \frac{\ell_{i}(j)}{D(j)}.
\end{align}
Our result is:
\begin{equation}
  C(t) \simeq \frac{6r\beta^3}{(1+\beta)^{2}(1 + \beta \omega)(1 - \beta \omega)^2} \frac{\ln t}{t}.
\end{equation}

When $\omega = -1 + \epsilon$, $\epsilon<< 1$,  we have $k_{max}= r+1, p_{k_{\ast}} \simeq \sqrt{\epsilon}$ and $\beta \simeq 1 - \sqrt{\epsilon}$. Thus, the expression for $C(t)$ becomes:
%\begin{eqnarray}
%  C(t) &\simeq& \frac{6 (r \phi)^3}{r(1 + \phi)\{r(1 + \phi) - 1\}(1 + \phi_{\omega})(1 - %%\phi_{\omega})^2} \frac{\ln t}{t} \nonumber \\
 % &\simeq& \frac{6 r^3}{(2r)^2 \sqrt{\epsilon}^2} \frac{\ln t}{t} \nonumber \\
%  &\simeq& \frac{3r}{8 \sqrt{\epsilon}} \frac{\ln t}{t}
%\end{eqnarray}
\begin{equation}
    C(t) \simeq \frac{3r}{8\sqrt{\epsilon}}\frac{\ln t}{t}.
\end{equation}
We obtain:
\[
\sqrt{\epsilon} C(t) \simeq \frac{3r}{8} \frac{\ln t}{t}
\]
Finally, when $\varepsilon\sim (\ln t/t)^2$, we have:
\begin{equation}
  C(t) \sim \frac{3r}{8} \sim \text{constant}.
\end{equation}

\begin{figure}[htbp]
  \centering
  \begin{subfigure}[b]{0.48\textwidth}
    \centering
    \includegraphics[width=\textwidth]{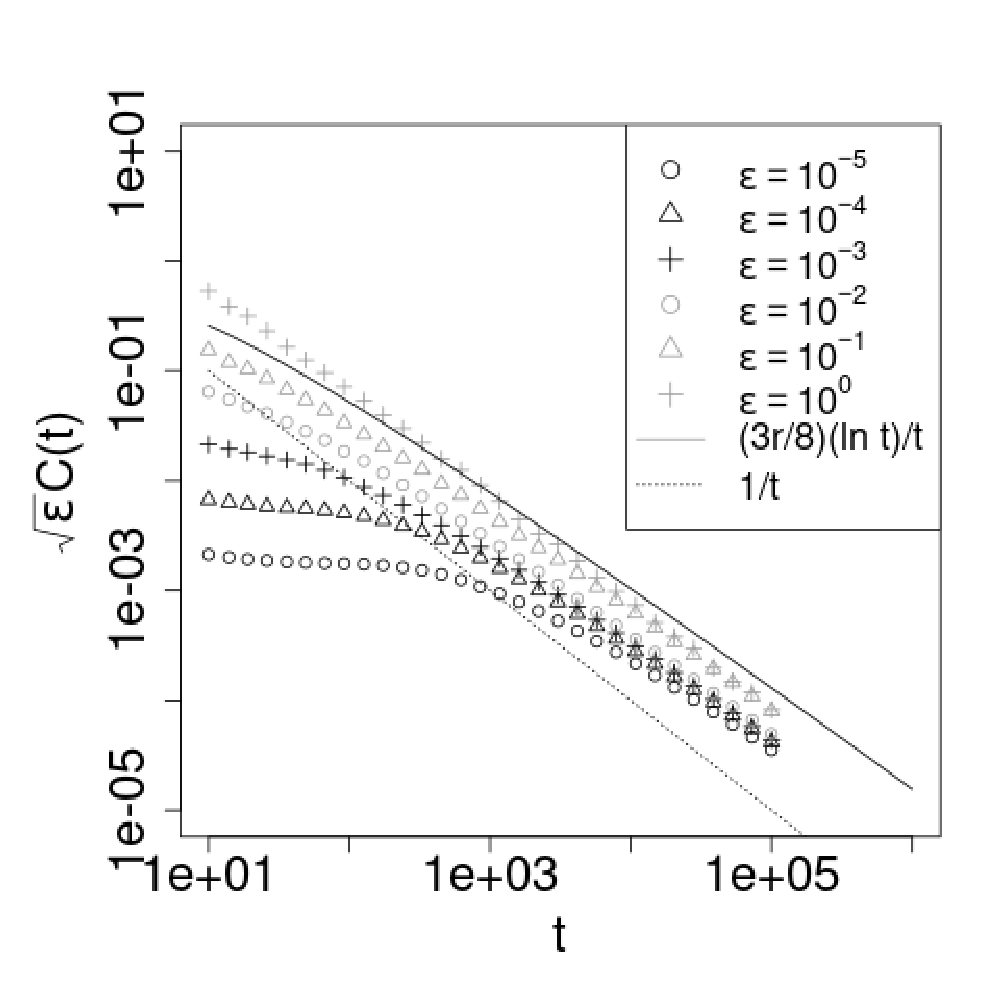}
  \end{subfigure}
  \hfill
  \begin{subfigure}[b]{0.48\textwidth}
    \centering
    \includegraphics[width=\textwidth]{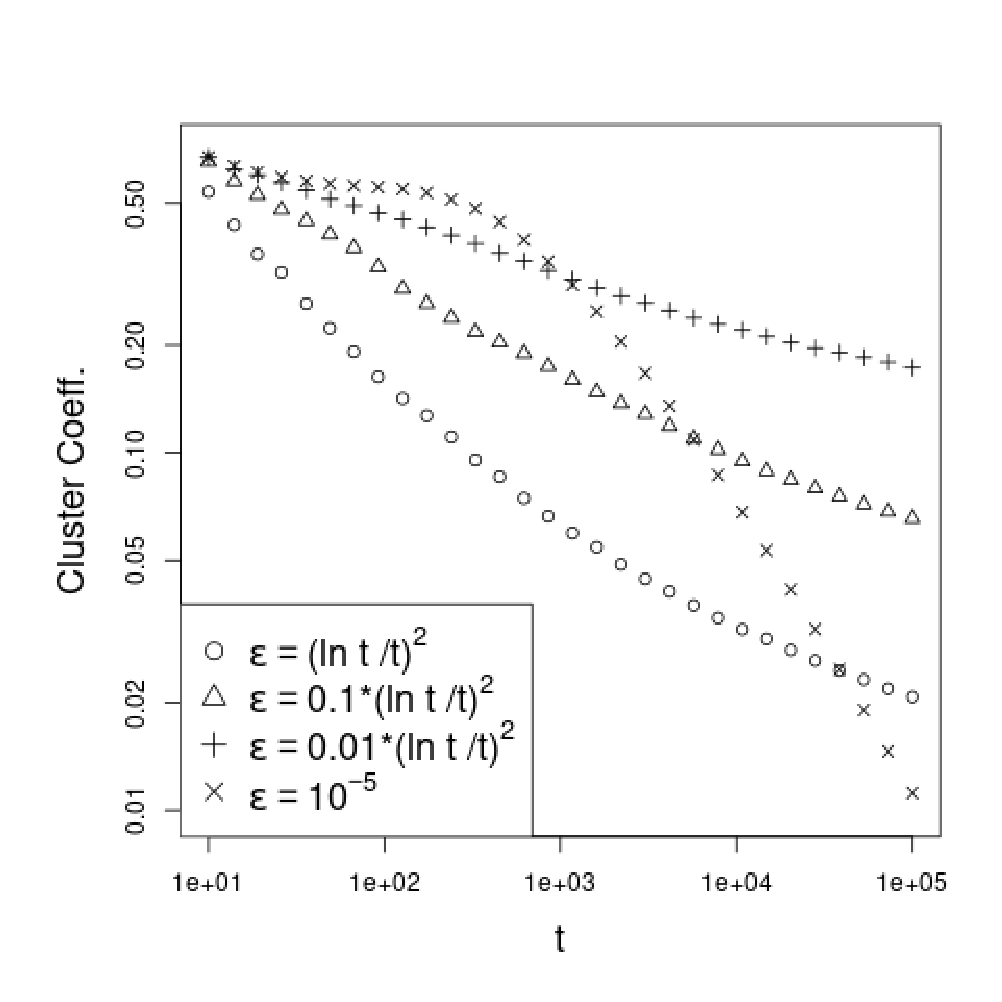}
  \end{subfigure}
  \caption{Plots of the clustering coefficient vs. $t$. In both plots, we use $N=10^5$, $r=3$, and 10 samples. In the left plot, we show the cases of  $\omega = -1 + \epsilon$ and $\varepsilon\in \{10^{-5}, 10^{-4}, \dots, 10^0\}$. The solid line shows the theoretical result
  $C(t)\sim \frac{3r}{8} \frac{\ln t}{t}$.
  In the right plot, the cases of $\omega = -1 + (\ln t/t)^2$ and 
   $\omega = -1 + a 10^{-5},a\in \{1,0.1,0.01\}$ are shown.}
  \label{fig:fig8}
\end{figure}

Fig.~\ref{fig:fig8} shows the plots of $C(t)$ vs. $t$.
In the left plot, $\varepsilon \in \{10^{-5}, \dots, 10^{0}\}$.
As can be clearly seen, $C(t)$ asymptotically converges to zero. 
By comparing the slopes of the MC data with the slope of $\ln t/t$, 
$C(t)$ behaves as $C(t) \sim \ln t / t$ for large $t$.
In the right plot, $\varepsilon = a(\ln t/t)^2,a\in \{1,0.1,0.01\}$ and $10^{-5}$ are considered.
The decay of $C(t)$ slows down as $t$ increases in the former case.
It is difficult to determine whether $C(t)$ converges to a positive constant 
in the large $t$ limit in the case.

\subsection{Mean end-to-end distance}

We study the mean end-to-end distance of the networks numerically.
We adopt $r=3,N=10^5$ and collnect $10^1$ samples.

\begin{figure}[htbp]
  \begin{subfigure}[b]{0.48\textwidth}
    \centering
    \includegraphics[width=\textwidth]{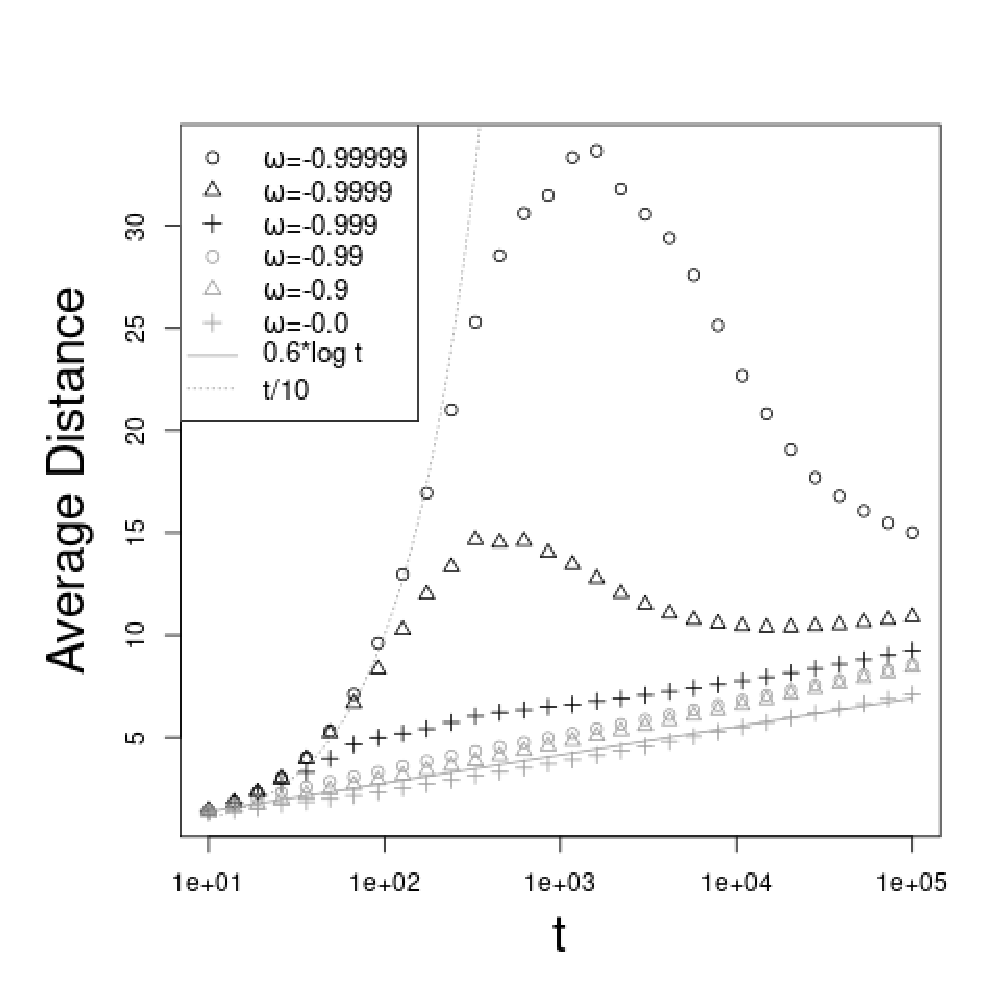}
  \end{subfigure}
  \hfill
  \begin{subfigure}[b]{0.48\textwidth}
    \centering
    \includegraphics[width=\textwidth]{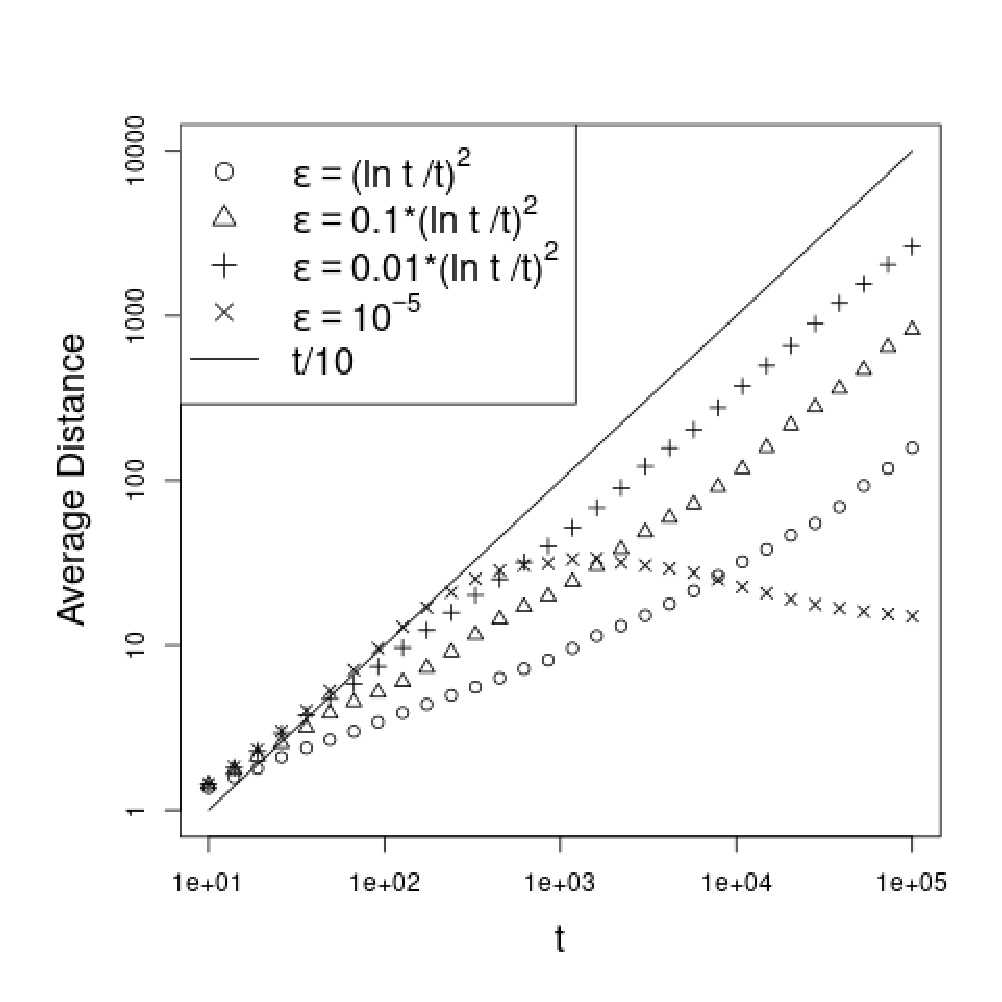}
  \end{subfigure}
  \caption{
  Plots of the 
  mean end-to-end distance vs. $t$.
  In both plots, we use $N=10^5$, $r=3$, and 10 samples.
  In the left plot, we show the cases of  $\omega = -1 + \epsilon$ and 
  $\varepsilon\in \{10^{-5}, 10^{-4}, \dots, 10^0\}$.
  The solid (broken) line shows the plot of $0.6\ln t (t/10)$.
  In the right plot, the cases of $\omega = -1 + (\ln t/t)^2$ and $\omega = -1 + a 10^{-5},a\in \{1,0.1,0.01\}$ are shown. The solid line shows the plot of $t/10$.
  }
  \label{fig:fig9}
\end{figure}

Fig.~\ref{fig:fig9} shows the plots of the mean end-to-end distance vs. $t$.
In the left plot, for the cases $\varepsilon \in \{10^{-4}, 10^{-5}\}$, 
one observes a crossover between the lattice and the random graph behaviors. 
For small $t$, the average distance is 
proportional to $t$, which is the expected behavior for a lattice.
For large $t$, the average distance becomes proportional to $\ln t$, 
which is characteristic of a random network. For other cases, only a 
logarithmic increase in the average distance is observed.
The right plot shows the average distance vs. $t$ for 
the cases $\varepsilon = a(\ln t/t)^2,a\in \{1,0.1,0.01\}$ and $10^{-5}$.
In the case $\varepsilon = (\ln t/t)^2$, the average distance 
increases linearly with $t$. While the study of the clustering coefficient
suggests the possibility that $C(t)$ converges to a positive value,
the average distance does not behave like $\ln t$, indicating that 
the network does not exhibit the small-world property.

We can understand why the mean end-to-end distance follows $\sim \ln t$ for $\omega = -1 + \epsilon$ with $\varepsilon\ll 1$ through a simple argument. In the construction of small-world networks \cite{WA1998}, the rewiring mechanism was proposed. When the probability of rewiring is positive, the mean end-to-end distance follows $\sim \ln t$. In the asymmetric BA model, the rewiring probability is proportional to $\sqrt{\varepsilon}$ in the stationary state
and is positive  for $\varepsilon > 0$. When $\varepsilon$ is too small, it is necessary
to wait a long time to reach the stationary state. Up to the time, the mean end-to-end 
distance follows $\sim  t$. 
One observe the cross-over behavior in the left figure of 
Fig.~\ref{fig:fig9}. When $\varepsilon=10^{-5}$, the mean end-to-end distance obeys
$\propto t$ for $t <10^2$.

\section{\label{sec:conclusion}Conclusion}
In this study, we analyzed in detail how changes in the parameter $\omega$ in the asymmetric Barabási-Albert (BA) model affect the structure of the network. While it is already known that an extended lattice is realized at $\omega = -1$ and a random graph at $\omega = 0$, this study focused on the intermediate region $-1 < \omega < 0$ and gained important insights into the small-world property and clustering characteristics of the network.

The main results of this study are as follows:

\begin{itemize}
    \item \textbf{Degree distribution for $\omega = -r/k,k\in\{r,r+1,\cdots\}$}: We derived the exact degree distribution for $\omega = -r/k$. Specifically, in the limit $k \to \infty$ (i.e., as $\omega \to 0$), we showed that the degree distribution follows a geometric distribution.

    \item \textbf{Perturbative calculation of the degree distribution for $\omega = -r/k + \varepsilon,k\in \{r,r+1,\cdots\}$}: We performed a perturbative calculation of the degree distribution for $\omega = -r/k + \varepsilon$. In particular, for $\omega = -1 + \varepsilon$, it was found that the first-order term is proportional to $\sqrt{\varepsilon}$.

    \item \textbf{Asymptotic behavior of the clustering coefficient and absence of small-world properties}: We analyzed the asymptotic behavior of the clustering coefficient $C(t)$, showing that it decreases proportionally to $\ln t/t$. Even with a slight change from $\omega = -1$ towards $\omega=-1+\epsilon,\epsilon<<1$, the network does not exhibit small-world properties.
\end{itemize}

Overall, this study provides a detailed understanding of how the asymmetric BA model influences network structure, particularly in terms of degree distribution, clustering coefficient, and the presence or absence of small-world properties. The results highlight the flexibility of the model in reproducing a wide variety of network structures, making it a valuable tool for modeling real-world systems such as social networks, biological networks, and infrastructure.

Future research could explore the dynamics of network structure when the parameter $\omega$ changes over time. The model could be generalized to include effects similar to pheromone evaporation in social insects like ants, potentially leading to steady-state network structures. Previous studies in symmetric BA models with $\omega = 1$ have already examined such phenomena \cite{Kim2006,Tadic2004}. Extending 
these works to the asymmetric model may provide further insights. Additionally, investigating the relationship between probabilistic models on these networks and the resulting network structures remains an important area for future research.

%\bibliography{myref202109}% Produces the bibliography via BibTeX.

\appendix

\section{Distribution of out-degree with $\omega = -1$}
\label{sec:omega-1}

We calculate $\{k_{i}^{\OUT}(t)\}$ for $r = 4$ and summarize it in Table \ref{tab:out-degrees}. We observe that the time evolution is deterministic, and the equality $\sharp G_{-1}(t) = r$ holds for any $t$ greater than or equal to $r$.

\begin{table}[htbp]
  \centering
  \begin{tabular}[b]{|c|c|c|c|c|c|c|c|c|c||c|c|}\hline
    Node & 1 & 2 & 3 & 4 & 5 & 6 & 7 & 8 & 9 & $G_{-1}(t)$ & $\sharp G_{-1}(t)$ \\ \hline\hline
    $t=1$ & 0 & \multicolumn{8}{c||}{} & \{1\} & 1 \\ \cline{1-3}\cline{11-12}
    $t=2$ & 1 & 0 & \multicolumn{7}{c||}{} & \{1, 2\} & 2 \\ \cline{1-4}\cline{11-12}
    $t=3$ & 2 & 1 & 0 & \multicolumn{6}{c||}{} & \{1, 2, 3\} & 3 \\ \cline{1-5}\cline{11-12}
    $t=4$ & 3 & 2 & 1 & 0 & \multicolumn{5}{c||}{} & \{1, 2, 3, 4\} & 4 \\ \cline{1-6}\cline{11-12}
    $t=5$ & 4 & 3 & 2 & 1 & 0 & \multicolumn{4}{c||}{} & \{2, 3, 4, 5\} & 4 \\ \cline{1-7}\cline{11-12}
    $t=6$ & 4 & 4 & 3 & 2 & 1 & 0 & \multicolumn{3}{c||}{} & \{3, 4, 5, 6\} & 4 \\ \cline{1-8}\cline{11-12}
    $t=7$ & 4 & 4 & 4 & 3 & 2 & 1 & 0 & \multicolumn{2}{c||}{} & \{4, 5, 6, 7\} & 4 \\ \cline{1-9}\cline{11-12}
    $t=8$ & 4 & 4 & 4 & 4 & 3 & 2 & 1 & 0 & \multicolumn{1}{c||}{} & \{5, 6, 7, 8\} & 4 \\ \hline
    $t=9$ & 4 & 4 & 4 & 4 & 4 & 3 & 2 & 1 & 0 & \{6, 7, 8, 9\} & 4 \\ \hline
  \end{tabular}
  \caption{Time evolution of out-degrees with $\omega = -1$ and $r = 4$. Blank spaces correspond to nodes that do not exist.}
  \label{tab:out-degrees}
\end{table}

In general, we obtain the following result:
\begin{equation}
  \label{eq:omega-1}
  \begin{split}
    &k_{i}^{\OUT}(t) = \min\{t-i, r\}, \quad i = 1, \cdots, r + s, \\
    &G_{-1}(t) = \{t, t-1, \cdots, \max(1, t-r+1)\}, \\
    &\sharp G_{-1}(t) = \min(t, r).
  \end{split}
\end{equation}

\section{Inequality $\sharp G_{\omega}(t) \ge r$ for $t \ge r$}
\label{sec:proofofG}

We shall show that $\sharp G_{\omega}(t) \ge r$ when $t \ge r$. From the definitions of $G_{\omega}(t)$ in eq.~\eqref{eq:Gomegat} and $\ell_{i}(t)$ in eq.~\eqref{eq:popularity}, it is clear that $G_{\omega}(t) \supset G_{\omega^{\prime}}(t)$ if $\omega \ge \omega^{\prime}$. Therefore,
\begin{equation}
  \label{eq:proofofG}
  \sharp G_{\omega}(t) \ge \sharp G_{\omega^{\prime}}(t), \quad
  \text{if} \quad
  \omega \ge \omega^{\prime}.
\end{equation}
Since $\omega \ge -1$, eqs.~\eqref{eq:omega-1} and \eqref{eq:proofofG} imply that $\sharp G_{\omega}(t) \ge \min(t, r)$. Our proposition immediately follows.

\section{Function $\chi(x, n)$}
\label{sec:function-chi}

For a positive integer $n$ and a positive real number $x$, the function $\chi(x, n)$ is defined as:
\begin{equation}
  \chi(x, n) = \sum_{k=0}^{n-1} \frac{1}{x + k}.
\end{equation}
In general, $\chi(z, w)$ can be expressed as:
\begin{equation}
  \chi(z, w) = \psi(z + w) - \psi(z),
\end{equation}
where $z$ and $w$ are complex numbers, and $\psi(z)$ is the digamma function \cite{AS}:
\begin{equation}
  \psi(z) = \frac{d}{dz} \ln \Gamma(z).
\end{equation}

\section{Why does the $\epsilon^{1/2}$ term appear?}
\label{sec:expansion}

We expand equation \eqref{eq:qk*} in powers of
$\epsilon$ for cases 3 and 4 as follows:
\begin{equation}
  \varepsilon b_{0} + \epsilon^{2}c_{0}
    + (a_{1} + \varepsilon b_{1})x
    + a_{2}x^{2}
    + O(x^{3}) = 0,
\end{equation}
where
\begin{equation}
    x = \left(\frac{r}{k_{\ast}} + \varepsilon k_{\max}\right)p_{k_{\ast}}, \quad
    a_{1} = \frac{(k_{\ast}-1)!r}{B(k_{\ast}/r-1,k_{\max})}.
\end{equation}
The remaining coefficients $b_{0}, c_{0}, b_{1}, a_{2}$
take some non-zero values.
We seek a solution that converges to zero as
$\epsilon$ approaches $0$,
because $I_{k_{\max}}$ defined in \eqref{eq:Ikt}
approaches the empty set as $\varepsilon\to 0$.
In case 3, $a_{1}$ is positive since $k_{\ast} > r$.
The lowest-order term of the solution is given by:
\begin{equation}
    x = \epsilon(-b_{0}/a_{1}) + \cdots.
\end{equation}
However, in case 4, we have $a_{1} = 0$.
Therefore, the solution becomes:
\begin{equation}
    x = \epsilon^{1/2}(-b_{0}/a_{2})^{1/2} + \cdots.
\end{equation}

\end{document}